\documentclass[preprint,showpacs,amsmath,amssymb]{revtex4}

\usepackage{graphicx}
\usepackage{latexsym}
\usepackage{amsmath}
\usepackage{amssymb}
\usepackage{amsfonts}
\usepackage{bm}

\newcommand{\la}{\left<}
\newcommand{\ra}{\right>}

\newcommand{\rvec}{\ensuremath{\bm{r}}}
\newcommand{\lvec}{\ensuremath{\bm{l}}}
\newcommand{\uvec}{\ensuremath{\bm{u}}}
\newcommand{\vvec}{\ensuremath{\bm{v}}}

\newcommand{\kB}{\mbox{$k_{\rm B}$}}
\newcommand{\kBT}{\mbox{$k_{\rm B}T$}}
\newcommand{\kappaT}{\mbox{$\kappa_{\rm T}$}}
\newcommand{\gT}{\mbox{$g$}}

\newcommand{\rN}{\ensuremath{r_\mathrm{N}}}
\newcommand{\rs}{\ensuremath{r_\mathrm{s}}}
\newcommand{\RN}{\ensuremath{R_\mathrm{N}}}
\newcommand{\Rs}{\ensuremath{R_\mathrm{s}}}
\newcommand{\Ps}{\ensuremath{P_\mathrm{s}}}

\newcommand{\be}{\ensuremath{b}}

\newcommand{\rhostar}{\ensuremath{\rho^*}}
\newcommand{\Ustar}{\ensuremath{U^*}}
\newcommand{\overlap}{\ensuremath{\varepsilon}}

\newcommand{\MSDall}{\ensuremath{h}}
\newcommand{\MSDmon}{\ensuremath{h}}
\newcommand{\MSDcmN}{\ensuremath{h_\mathrm{N}}}
\newcommand{\MSDcms}{\ensuremath{h_\mathrm{s}}}

\newcommand{\DN}{\ensuremath{D_\mathrm{N}}}
\newcommand{\Ds}{\ensuremath{D_\mathrm{s}}}
\newcommand{\Ts}{\ensuremath{T_\mathrm{s}}}
\newcommand{\TN}{\ensuremath{T_\mathrm{N}}}
\newcommand{\Te}{\ensuremath{T_\mathrm{e}}}
\newcommand{\de}{\ensuremath{d_\mathrm{e}}}
\newcommand{\CN}{\ensuremath{C_\mathrm{N}}}
\newcommand{\CNstar}{\ensuremath{C_\mathrm{T_N}}}
\newcommand{\Cs}{\ensuremath{C_\mathrm{s}}}
\newcommand{\Csstar}{\ensuremath{C_\mathrm{T_s}}}
\newcommand{\tstar}{\ensuremath{t^*}}

\newcommand{\pjump}{\ensuremath{\bm{u}(0)}}

\newcommand{\deltat}{\ensuremath{\delta t}}

\begin{document}

\title{Scale-free center-of-mass displacement correlations in polymer melts
       without topological constraints and momentum conservation:
       A bond-fluctuation model study}

\author{J.P.~Wittmer}
\email{joachim.wittmer@ics-cnrs.unistra.fr}
\affiliation{Institut Charles Sadron, Universit\'e de Strasbourg, CNRS, 23 rue du Loess, 67037 Strasbourg Cedex, France}
\author{P.~Poli\'nska}
\affiliation{Institut Charles Sadron, Universit\'e de Strasbourg, CNRS, 23 rue du Loess, 67037 Strasbourg Cedex, France}
\author{A.~Cavallo}
\affiliation{Dipartimento di Fisica, Universit\`a degli Studi di Salerno, via Ponte don Melillo, I-84084 Fisciano, Italy}
\author{H.~Meyer}
\affiliation{Institut Charles Sadron, Universit\'e de Strasbourg, CNRS, 23 rue du Loess, 67037 Strasbourg Cedex, France}
\author{J.~Farago}
\affiliation{Institut Charles Sadron, Universit\'e de Strasbourg, CNRS, 23 rue du Loess, 67037 Strasbourg Cedex, France}
\author{A.~Johner}
\affiliation{Institut Charles Sadron, Universit\'e de Strasbourg, CNRS, 23 rue du Loess, 67037 Strasbourg Cedex, France}
\author{J.~Baschnagel}
\affiliation{Institut Charles Sadron, Universit\'e de Strasbourg, CNRS, 23 rue du Loess, 67037 Strasbourg Cedex, France}

\date{\today}

\begin{abstract}
By Monte Carlo simulations of a variant of the bond-fluctuation model without topological constraints 
we examine the center-of-mass (COM) dynamics of polymer melts in $d=3$ dimensions. Our analysis focuses 
on the COM displacement correlation function $\CN(t) \approx \partial_t^2 \MSDcmN(t)/2$, 
measuring the curvature of the COM mean-square displacement $\MSDcmN(t)$. We demonstrate that 
$\CN(t) \approx -(\RN/\TN)^2 (\rhostar/\rho) \ f(x=t/\TN)$ with $N$ being the chain length ($16 \le N \le 8192$),
$\RN\sim N^{1/2}$ the typical chain size,
$\TN\sim N^2$ the longest chain relaxation time,
$\rho$ the monomer density,
$\rhostar \approx N/\RN^d$ the self-density and
$f(x)$ a universal function decaying asymptotically as
$f(x) \sim x^{-\omega}$ with $\omega = (d+2) \times \alpha$
where $\alpha = 1/4$ for $x \ll 1$ and $\alpha = 1/2$ for $x \gg 1$.
We argue that the algebraic decay $N \CN(t) \sim - t^{-5/4}$ for $t \ll \TN$ results from an interplay 
of chain connectivity and melt incompressibility giving rise to the correlated motion of chains and subchains. 
\end{abstract}

\maketitle

%%%%%%%%%%%%%%%%%%%%%%%%%%%%%%%%%%%%%%%%%%%
\section{Introduction}
\label{sec_intro}

\paragraph*{Prelude: Overdamped colloidal suspensions.}
\label{par_intro_colloid}
Dense, essentially incompressible simple liquids with conserved momentum  are known to 
exhibit long-range correlations of the particle displacement field \cite{HansenBook,DhontBook,Alder70}. 
As first shown in the seminal molecular dynamics (MD) simulations by Alder and Wainwright \cite{Alder70},
the coupling of displacement and momentum fields manifests itself by an algebraic decay 
of the velocity correlation function (VCF), 
\begin{equation}
C(t) \equiv \la \vvec(t) \cdot \vvec(0) \ra 
\sim + 1/\xi(t)^{d} \sim +1/t^{\alpha d},
\label{eq_Alder}
\end{equation}
with $\vvec(t)$ being the particle velocity at time $t$, $d$ the spatial dimension
and $\xi(t) \sim t^{\alpha}$ the typical particle displacement with exponent 
$\alpha = 1/2$ \cite{foot_Alder}. 
\begin{figure}[t]
\centerline{\resizebox{0.6\columnwidth}{!}{\includegraphics*{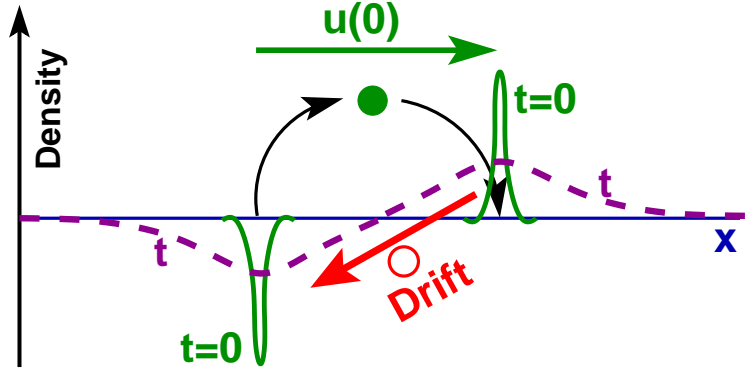}}}
\caption{The displacement auto-correlation function $C(t)$ of dense colloids without momentum 
conservation is known to reveal a negative algebraic tail due to long-range dynamical interactions 
caused by the weak compressibility of the solution \cite{DhontBook}: 
If a particle (filled sphere) is displaced at time $t=0$ by a distance $u(0)$
along the $x$-axis, this creates a density dipole (bold peaks) decaying in time
by the cooperative diffusion of the density field (dashed line).
The gradient of the density field generates a force pulling
the particle (open sphere) back to its original position. 
We argue that a related mechanism is relevant for describing the 
correlations of the chain and subchain center-of-mass displacements 
in dense polymer melts without topological constraints and momentum conservation. 
\label{fig_dipole}
}
\end{figure}
Interestingly, even if the momentum conservation is dropped,
as justified for overdamped dense colloidal suspensions \cite{DhontBook},
scale-free albeit much weaker correlations of the displacement field are to be expected
due to the incompressibility constraint \cite{DhontBook}. 
As illustrated in Fig.~\ref{fig_dipole}, the motion of a tagged colloid is coupled to the 
collective density dipole field \cite{foot_colloid}, 
\begin{equation}
\delta \rho(\rvec,t) \approx \frac{1}{\xi^d(t)} \frac{\pjump \cdot \rvec}{\xi^2(t)} \ e^{-(\rvec/\xi(t))^2} 
\mbox{ for } t > 0,
\label{eq_dipolefield}
\end{equation}
created by the colloid's own displacement $\pjump$ at $t=0$.
After averaging over the typical displacements of the test particle and assuming a 
Cahn-Hilliard response proportional to the gradient $\nabla \mu(\rvec)$ of the chemical potential $\mu(\rvec)$
of the density field, this leads to a {\em negative} algebraic long-time decay of the VCF %\cite{foot_colloid}
\begin{equation}
C(t) \sim -1/\xi(t)^{d+2} \sim - 1/t^{\omega}
\mbox{ with } \omega = (d+2) \ \alpha,
\label{eq_VCFcolloid}
\end{equation}
e.g., $\omega=5/2$ in $d=3$ dimensions.
This phenomenological scaling picture agrees with mode-coupling calculations 
\cite{DhontBook,fuchs02,hagen97}
and has been confirmed computationally by means of Lattice-Boltzmann simulations \cite{hagen97},
MD simulations \cite{williams06} and even Monte Carlo (MC) simulations with local moves (as discussed below) 
\cite{foot_MC}.
\paragraph*{Deviations from Flory's ideality hypothesis.}
\label{par_intro_static}
In this study we explore the dynamics of another complex fluid for which momentum conservation is generally believed 
to be irrelevant \cite{DoiEdwardsBook}: melts of long and flexible homopolymers of length $N$ in $d=3$ dimensions.
Following Flory's ``ideality hypothesis" \cite{DegennesBook,DoiEdwardsBook}, one expects these chains to obey a 
Gaussian statistics with a typical chain size $\RN \approx b N^{1/2}$ where $b$ stands for the effective bond 
length of asymptotically long chains \cite{DoiEdwardsBook}. 
Recently, this cornerstone of polymer physics has been challenged both theoretically and numerically 
for three-dimensional (3D) melts \cite{WMBJOMMS04,WBM07,WCK09}, 
for effectively two-dimensional (2D) ultrathin films \cite{ANS03,CMWJB05}
and one-dimensional (1D) thin capillaries \cite{Brochard79,LFM11}.
The physical idea behind the predicted long-range correlations is related to the 
``segmental correlation hole", $\rhostar(s) \approx s/\Rs^d$, of a subchain of arc-length $s$ 
of typical size $\Rs$ \cite{WBM07}.
Due to the overall incompressibility of the melt this sets an entropic penalty
$\Ustar(s) \approx \rhostar(s)/\rho$ (with $\rho$ being the total monomer density)
against bringing two subchains together \cite{DegennesBook,ANS03,WBM07}.
In $d=3$ dimensions, the segmental correlation hole effect is weak, $\Ustar(s) \sim 1/\sqrt{s}$, 
and a perturbation calculation can be performed \cite{WMBJOMMS04,WBM07}.
The detailed calculation yields, e.g., for the intrachain angular correlations function $\Ps$ \cite{foot_Psdef}
an algebraic decay \cite{WMBJOMMS04,WBM07,WCK09},
\begin{equation}
\Ps = \sqrt{\frac{3}{8\pi^3}} \frac{1}{b^3 \rho} \frac{1}{s^{3/2}}
\mbox{ for } \gT \ll s \ll N,
\label{eq_PsD3}
\end{equation}
at variance to Flory's hypothesis with $\gT$ being the subchain length spanning the screening length
of the density fluctuations \cite{DoiEdwardsBook}. 
Since $\Ps \sim \partial_s^2\Rs^2$, the angular correlation function provides a direct measure of the 
curvature of the mean-squared subchain size $\Rs^2$ and
Eq.~(\ref{eq_PsD3}) implies that $1- \Rs^2/b^2s \approx \Ustar(s) \sim 1/\sqrt{s}$ \cite{WMBJOMMS04,WBM07}.
Interestingly, $\gT$ is related to the isothermal compressibility $\kappaT$ of the solution
\cite{ANS03,ANS05a,WCK09} and can thus be determined directly from the total monomer structure factor
\begin{equation}
G(q)
\equiv \frac{1}{\rho V} \sum_{n,m=1}^{\rho V} 
\la e^{- i \bm{q} \cdot (\bm{r}_n - \bm{r}_m) }\ra
\stackrel{q\to 0}{\stackrel{N\to \infty}{\Longrightarrow}} \gT
\equiv  T \ \kappaT \rho,
\label{eq_gdef}
\end{equation}
$\bm{r}_n$ being the position of monomer $n$,
$\bm{q}$ the conjugated wavevector,
$V = L^d$ the volume of the system
and $T$ the temperature.
(Boltzmann's constant $\kB$ \ is set to unity throughout this paper.)
Due to its definition, $\gT$ is often called ``dimensionless compressibility" \cite{WBM07}.
Remarkably,  Eq.~(\ref{eq_PsD3}) does no depend {\em explicitly} on $\gT$.
This reflects the fact that the deviations arise due to the incompressibility
of ``blobs" \cite{DegennesBook} on scales corresponding to $s \gg \gT$ 
\cite{WMBJOMMS04,WCK09}.
\paragraph*{Aim of this study.}
\label{par_intro_aim}
Naturally, Eq.~(\ref{eq_PsD3}) and related findings beg the question of whether a similar interplay 
between the connectivity of the chains and the incompressibility of the melt may cause measurable 
scale-free and $\gT$-independent {\em dynamical} correlations between chains and between subchains. 
To avoid additional physics and to simplify the problem we focus on polymer melts where hydrodynamic 
\cite{ANS11a} and topological constraints may be considered to be negligible,
as in the pioneering work by Paul {\em et al.} \cite{Paul91a,Paul91b},
or are deliberately switched off \cite{Shaffer94,Shaffer95,WBM07,WCK09}. 
Deviations from the expected Rouse-type dynamics \cite{DoiEdwardsBook} have indeed been reported 
for such systems in various numerical
\cite{Paul91a,Paul91b,Shaffer95,KBMB01,HT03,Briels02}
and experimental studies \cite{Paul98,Smith00,PaulGlenn}.
Characterizing, e.g., the motion of the chain center-of-mass (COM) $\rN(t)$ by its 
mean-square-displacement (MSD) $\MSDcmN(t)$,  it was found that 
\begin{equation}
\MSDcmN(t) \equiv \la ( \rN(t) - \rN(0) )^2 \ra \sim t^{\beta} 
\mbox{ for } t \ll \TN
\label{eq_betadef}
\end{equation}
with $\TN \sim N^2$ being the longest (Rouse) relaxation time and $\beta \approx 0.8$ an 
empirical exponent. This is of course at variance to the key assumption of the Rouse model
that the random forces acting on the monomers (and thus on the chain) are uncorrelated, 
which implies an exponent $\beta=1$ for all times \cite{DoiEdwardsBook}.
In this paper we attempt to clarify this problem by means of MC simulations of 
a variant of the bond-fluctuation model (BFM) \cite{BFM} 
using local hopping moves which do {\em not} conserve topology \cite{WBM07}
and assuming a finite monomer overlap penalty $\overlap$ \cite{WCK09}.
In analogy to our work on the angular correlation function $\Ps \sim \partial_s^2 \Rs^2$ \cite{WMBJOMMS04},
our analysis focuses on the COM displacement correlation function $\CN(t) \sim \partial_t^2 \MSDcmN(t)$, 
measuring the curvature of $\MSDcmN(t)$ \cite{foot_MC}. 
This allows us to probe directly the colored forces from the molecular bath \cite{DhontBook} 
acting on a tagged reference chain displaced at $t=0$ and, in turn, to elaborate the scaling theory 
sketched below.

\paragraph*{Key results.}
\label{par_intro_key}
We demonstrate numerically that the VCF $\CN(t)$ does indeed {\em not} vanish as it would 
if all the random forces acting on the chains {\em were} uncorrelated \cite{DoiEdwardsBook}. 
Instead, $\CN(t)$ is found to scale as 
\begin{equation}
\CN(t) \approx - \left(\frac{\RN}{\TN}\right)^2 \frac{\rhostar(N)}{\rho} \ f(t/\TN)
\label{eq_key} 
\end{equation} 
with $f(x)$ being a universal scaling function. 
Note that the postulated Eq.~(\ref{eq_key}) does not depend explicitly on the compressibility of the solution.
The squared characteristic ``velocity" $(\RN/\TN)^2$ arises for dimensional reasons.
The prefactor $\rhostar(N)/\rho$ is motivated by the correlation hole penalty, 
i.e. the incompressibility constraint 
which ultimately couples the displacements of (sub)chains \cite{foot_rhostartho}.
As one expects from Eq.~(\ref{eq_VCFcolloid}), the scaling function decays as $f(x) \sim 1/x^{\omega}$ 
with an exponent $\omega = (d+2)/2$ for $x \gg 1$. We show that this long-time behavior is preceded by 
a much weaker algebraic decay with an exponent \cite{foot_capillary}
\begin{equation}
\omega= (d+2) \alpha = 5/4
\mbox{ for } x \ll 1
\label{eq_keyomega}
\end{equation}
due to the much slower relaxation, $\alpha = 1/4$, of the collective dipole field of subchains 
which was generated by the initial displacement of a tagged subchain at $t=0$. 
The gradient of the chemical potential $\nabla \mu(\rvec)$ pulling the reference subchain back to 
its original position is of course not only due to the density fluctuation of the subchain density 
field but also to the tensional forces along the chains caused by the displacement. 
(Subchain density fluctuations and tensions are coupled and associated both to a 
free energy fluctuation of order $\kBT s^0 N^0$.)
Since $\CN(t) \sim \partial_t^2 \MSDcmN(t)$, it follows from Eq.~(\ref{eq_keyomega}) that 
for sufficiently short times, such that the white forces acting on the chains are negligible,
we expect to find 
\begin{equation}
\beta = 2 - \omega = (6-d)/4 = 3/4, 
\label{eq_omega2beta}
\end{equation}
which is rather similar to the fit, Eq.~(\ref{eq_keyomega}), suggested in the literature.
See Refs.~\cite{Schweizer89} and \cite{Guenza02} for two closely related theoretical studies.
Interestingly, Schweizer's mode-coupling theory approach \cite{Schweizer89} is consistent 
with Eq.~(\ref{eq_omega2beta}).
We stress that the presented numerical study is necessarily incomplete since important dynamical correlations 
are expected to arise in more realistic models due to topological constraints and, even more importantly, 
due to not fully screened hydrodynamic interactions which have recently been shown to matter \cite{ANS11a}.

\paragraph*{Outline.}
\label{par_intro_outline}
The paper is organized as follows.
In Sec.~\ref{sec_algo} the numerical algorithm is introduced and some technical details are discussed.
Our computational results are presented in Sec.~\ref{sec_comp} where we focus on essentially 
incompressible melts (Sec.~\ref{sub_comp_jean}) but also discuss effects of finite excluded volume 
(Sec.~\ref{sub_comp_robust}).
Our results are summarized in Sec.~\ref{sub_conc_summary}.
The paper concludes in Sec.~\ref{sub_conc_outlook} with a comment on what we would expect
if topology conservation is switched on again.

%%%%%%%%%%%%%%%%%%%%%%%%%%%%%%%%%%%%%%%%%%%
\section{Some algorithmic details}
\label{sec_algo}

\subsection{The classical bond-fluctuation model}
\label{sub_algo_BFMclassic}
\begin{figure}[t]
\centerline{\resizebox{0.6\columnwidth}{!}{\includegraphics*{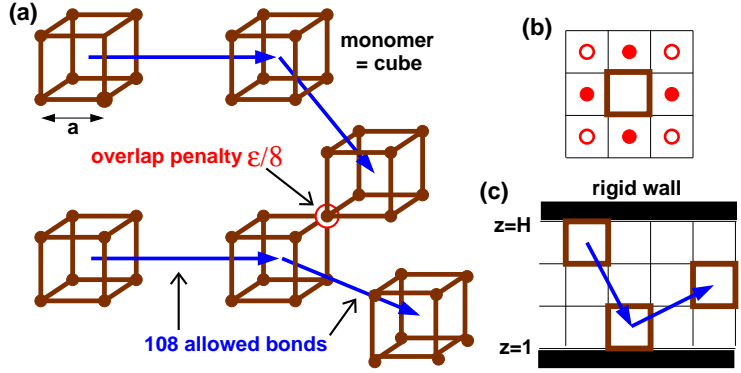}}}
\caption{
The BFM is a lattice MC algorithm for coarse-grained polymer chains where monomers are 
represented by cubes on a simple cubic lattice connected by a set of allowed bond vectors \cite{BFM}:
{\bf (a)}
We use here a variant of the BFM where a finite energy $\overlap$ has to be paid if two cubes totally overlap 
and a corresponding fraction for a partial monomer overlap \cite{WCK09}. 
{\bf (b)}
Using {\em local} MC jump attempts to the next (filled circles) and next-nearest (open circles)
neighbors we investigate the influence of the incompressibility constraint on the dynamics 
of polymer melts {\em without} topological constraints.
{\bf (c)} 
In Sec.~\ref{sub_comp_beads} we will consider beads ($N=1$) confined between hard walls 
\cite{CMWJB05,foot_capillary}.
\label{fig_algo}
}
\end{figure}

The classical BFM is an efficient lattice MC algorithm for coarse-grained polymer chains 
where each monomer occupies {\em exclusively} a unit cell of $2^d$ 
lattice sites on a $d$-dimensional simple cubic lattice \cite{BFM,Paul91a,Paul91b}.
(The fraction $\phi$ of occupied lattice sites is thus $\phi = 2^d \rho$.)
The BFM was proposed in 1988 by Carmesin and Kremer \cite{BFM} as an alternative to single-site 
self-avoiding walk models, which retains the computational efficiency of the lattice 
without being plagued by ergodicity problems. The key idea is to increase the size of 
the monomers and the number of bond vectors to allow a better representation of the 
continuous-space behavior of real polymer melts.
A widely used choice of bond vectors for the 3D variant of the BFM is given, e.g.,
by all the permutations and sign combinations of the six vectors \cite{Paul91a,Paul91b,KBMB01,HT03}
\begin{equation}
(2,0,0), (2,1,0), (2,1,1), (2,2,1), (3,0,0), (3,1,0).
\label{eq_bondset}
\end{equation}
If only {\em local} MC moves of the monomers to the six nearest neighbor sites are performed 
--- called ``L06" moves \cite{WBM07} --- this vector set ensures automatically that polymer chains cannot cross. 
(These L06-moves are represented in Fig.~\ref{fig_algo}(b) by the filled circles.)
Consequently, several authors report a reptation-type dynamics for chain lengths above $N \approx 200$ 
at a ``melt" volume fraction $\phi=8 \rho = 0.5$ \cite{Paul91a,Paul91b,KBMB01,HT03}.

\begin{table}[t]
\begin{tabular}{|c||c|c|c||c|c|c|}
\hline
$\overlap$ &   $\gT$ & $l$  &$\be$ & $A$   &  $W$  & $N \DN$  \\ \hline
0.0        & $\infty$&2.718 & 2.72 & 0.2109& 0.032 &0.065          \\
0.01       & 209     &2.718 & 2.80 & 0.2109& 0.030 &0.062           \\
0.1        & 22      &2.719 & 2.92 & 0.2067& 0.024 &0.058          \\
1          & 2.4     &2.721 & 3.13 & 0.1796& 0.015 &0.040       \\
10         & 0.32    &2.670 & 3.24 &8.8E-02& 0.003 &0.009          \\
100        & 0.25    &2.636 & 3.24 &6.9E-02& 0.0010&0.003          \\ 
\hline
\end{tabular}
\caption[]{Various properties for BFM polymers melts of arbitrarily large chain length $N$
at volume fraction $\phi = 8 \rho = 0.5$:
the dimensionless compressibility $\gT$,
the root-mean-square bond length $l$, 
the effective bond length $\be$,
the acceptance rate $A$,
the local monomer mobility $W$ and 
the self-diffusion coefficient $\DN$. 
The dynamical data refer to L26-moves to the nearest and next-nearest lattice sites. 
\label{tab_overlap}}
\end{table}

\subsection{BFM with topology violating local moves}
\label{sub_algo_BFMourapproach}
For consistency with these studies we keep Eq.~(\ref{eq_bondset}), although the non-crossing constraint 
is irrelevant for us, and use the same volume fraction $\phi=0.5$. With respect to the classical variant 
of the BFM in $d=3$ our algorithm differs in two important points:

{\em (i)}
As described in Ref.~\cite{WCK09}, we use a finite excluded volume penalty $\overlap$
which has to be paid if two monomers fully overlap. The overlap of two cube corners is sketched
in Fig.~\ref{fig_algo}(a). Temperature is arbitrarily set to unity.
Technically, the finite monomer interaction penalty is implemented using a Potts spin 
representation of the discrete local density field $\rho(\rvec)$ \cite{WCK09}.
The monomer overlap penalty parameter $\overlap$ is in fact a Laplace multiplier 
controlling the fluctuations of spins and imposing thus the 
incompressibility of the melt. By reducing the fluctuations of the spins 
the Laplace multiplier causes thus the effective entropic forces which imply 
the dynamical correlations discussed in Sec.~\ref{sec_comp}.
This modification allows us to check at one constant volume fraction ($\phi = 0.5$) 
that the dynamical correlations do not depend {\em explicitly} on the monomer excluded volume 
as stated in Eq.~(\ref{eq_key}). 

{\em (ii)}
We use local hopping moves to the 26 next and next-nearest lattice sites (so-called ``L26" moves).
Due to the larger jumps, chains cross and the dynamics is of Rouse-type:
the chain self-diffusion coefficient $\DN$, e.g., scales as $\DN \sim 1/N$
for all chain lengths $N$ even if monomer overlap is disallowed ($\overlap=\infty$) \cite{WBM07}.

Although the use of L26-moves and finite monomer interactions does speed up the relaxation dynamics, 
it remains obviously impossible to equilibrate dense polymer solutions with chain lengths up to $N=8192$ 
just using {\em local} hopping moves.
Taken advantage of our previous studies on static properties \cite{WBM07,WCK09},
we have used configurations equilibrated using a mix of {\em global} 
slithering snake and double bridging moves together with local L26-moves.
We use periodic simulation boxes of linear dimension $L=256$. 
Thus at $\phi=8 \rho = 0.5$ these systems contain $N M = \rho L^3 = 2^{20} \approx 10^6$ 
monomers and even for chains with $N=8192$ we still have $M=128$ chains.
\begin{figure}[tb]
\centerline{\resizebox{0.6\columnwidth}{!}{\includegraphics*{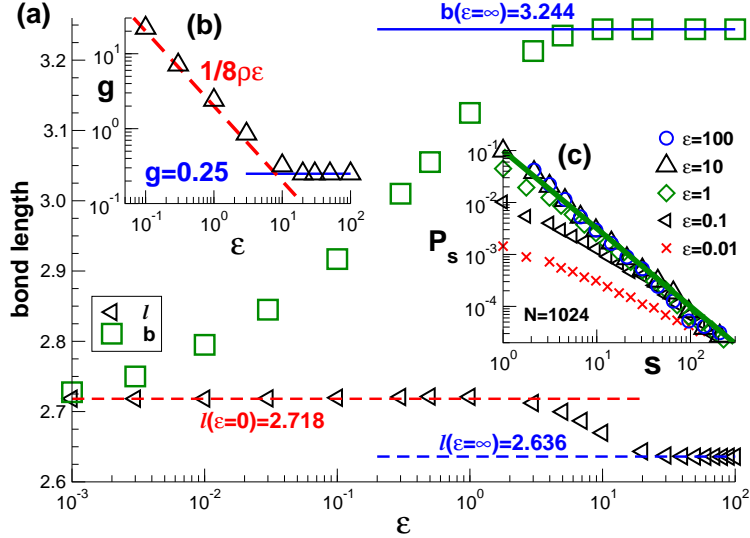}}}
\caption{Summary of static properties for BFM chains at volume fraction $\phi=8\rho=0.5$: 
{\bf (a)} 
Root mean-square bond length $l(\overlap)$ (triangles) and 
effective bond length $b(\overlap)$ (squares) in $d=3$ dimensions.
{\bf (b)}
The dimensionless compressibility $\gT(\overlap)$ for asymptotically long chains levels off
for large $\overlap$ where $\gT(\overlap \gg 10) \approx 0.25$ (bold line). 
{\bf (c)}
$\Ps$ for several $\overlap$ compared to the exponent $-3/2$ (bold line) 
expected for $s \gg \gT$ according to Eq.~(\ref{eq_PsD3}) \cite{WCK09}.
\label{fig_static}
}
\end{figure}

\begin{figure}[tb]
\centerline{\resizebox{0.6\columnwidth}{!}{\includegraphics*{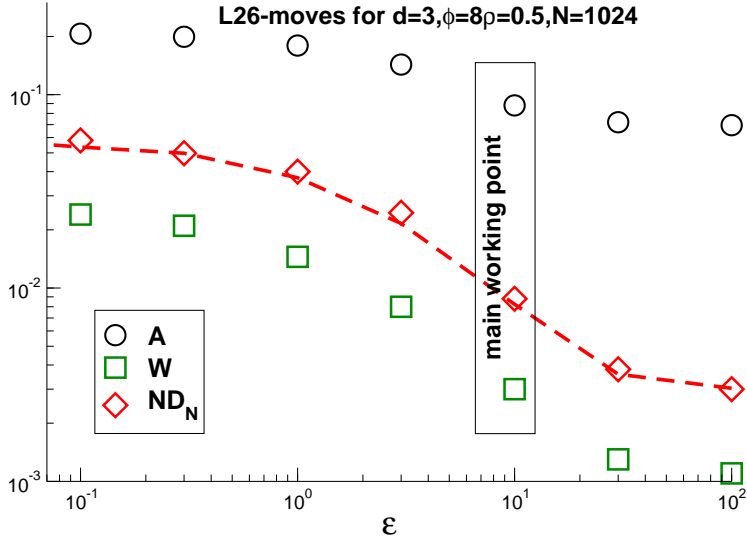}}}
\caption{Acceptance rate $A$, effective local mobility $W$ and self-diffusion coefficient $N \DN$
obtained using L26-moves at volume fraction $\phi=0.5$ {\em vs.} the monomer overlap penalty $\overlap$. 
All dynamical properties decrease monotonously with the interaction penalty but
become essentially constant above our main working point at $\overlap = 10$. 
The dashed line corresponds to the diffusion coefficient according 
to the Rouse prediction, Eq.~(\ref{eq_W2DN}), assuming the indicated mobilities $W$ (squares).
\label{fig_AW}
}
\end{figure}

Using these well characterized configurations we have computed time series
over $10^7$ or $10^8$ Monte Carlo Steps (MCS) using local L26-moves. We stress that the aim
is not to cover necessarily the full relaxation time $\TN \sim N^2$ for our larger chains,
but to precisely describe the dynamical correlations which are the most pronounced at short times.
As main working point we take $\overlap = 10$ for which the chain length has been  
scanned up to $N=8192$. % 
For other penalties $\overlap$ we present results obtained using $N=1024$.
Some relevant properties are summarized in Table~\ref{tab_overlap} and are represented in 
Fig.~\ref{fig_static} and Fig.~\ref{fig_AW}.

\subsection{Reminder of static properties}
\label{sub_algo_static}
Panel (a) of Fig.~\ref{fig_static} shows the mean-square bond length $l(\overlap)$
and the effective bond length $b(\overlap)$ for asymptotically long chains as determined 
in Ref.~\cite{WCK09}. Obviously, $b(\overlap) \to l(\overlap=0)=2.718$ in the small-$\overlap$ limit.
$b(\overlap)$ then increases in the intermediate $\overlap$-window before it levels off
at our main working point $\overlap = 10$. 
As one may expect, systems with $\overlap=100$ cannot be distinguished from systems
computed using the classical BFM without monomer overlap ($\overlap=\infty$).
Panel (b) presents the dimensionless compressibility $\gT(\overlap)$ 
obtained from total static structure factor, Eq.~(\ref{eq_gdef}).
The dashed line corresponds to the asymptotics for weak interactions, $1/\gT(\overlap) \sim \overlap$ \cite{WCK09}.
The compressibility levels off for large penalties where $\gT(\overlap \gg 10) \to 0.25$ (bold line).
The angular correlation function $\Ps$ \cite{foot_Psdef} is presented for several overlap penalties $\overlap$
and one chain length $N=1024$ in panel (c). The theoretical prediction Eq.~(\ref{eq_PsD3}) 
is nicely confirmed for $s \gg \gT(\overlap)$ by the power-law asymptote indicate by the bold line.
%

%%%%%%%%%%%%%%%%%%%%%%%%%%%%%%%%%%%%%%%%%%%
\section{Computational results}
\label{sec_comp}
\subsection{Diffusion of dense BFM beads}
\label{sub_comp_beads}
\begin{figure}
\centerline{\resizebox{0.6\columnwidth}{!}{\includegraphics*{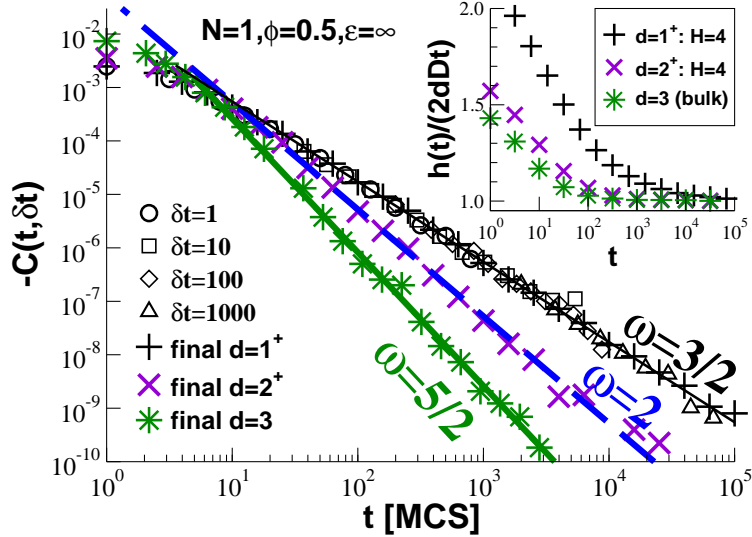}}}
\caption{Diffusion of BFM beads in $d=1^+$, $d=2^+$ and $d=3$ dimensions obtained using L26-moves.
Inset: Although the bead diffusion is essentially free, small deviations are visible
if $\MSDall(t)/(2d D t)$ is plotted in log-linear coordinates.
Main panel: Collapse of $C(t,\deltat)$ for $d=1^+$ and various time increments $\deltat$ (open symbols). 
The ``final" function $C(t)$ is obtained by adding the first decade of data for each $\deltat$
and logarithmic averaging. The cummulants agree nicely with the 
predicted exponent $\omega=(d+2)/2$, Eq.~(\ref{eq_VCFcolloid}).
\label{fig_BFMbeads}
}
\end{figure}

Since in MC simulations there is no ``mono\-mer mass",
no (conserved or non-conserved) ``monomer momentum" and not even an instantaneous velocity,
it might at first sight appear surprising that a well-posed ``velocity correlation function"
(VCF) can be defined and measured. To illustrate that this is indeed the case is the
first purpose of this subsection. The second is to verify that the negative analytic decay 
of the VCF expected for overdamped colloids, Eq.~(\ref{eq_VCFcolloid}),
is also of relevance for dense BFM beads ($N=1$) diffusing through configuration space 
by means of {\em local} hopping moves on the lattice.
The systems presented in Fig.~\ref{fig_BFMbeads} correspond to three different effective dimensions $d$.
The effectively 1D systems ($d=1^+$) have been obtained by confining the beads to a thin
capillary of square cross-section, the 2D systems ($d=2^+$) by confining the
beads to a thin slit as shown in Fig.~\ref{fig_algo}(c). 
The distance $H=4$ between parallel walls allows the free crossing of the beads. 
(For the 1D case this is crucial, of course.)
The data has been obtained for beads without monomer overlap ($\overlap=\infty$)
by means of L26-moves on a 3D cubic lattice for a volume fraction $\phi=0.5$.
We average over the $2^{20}$ beads contained in each configuration.
One standard measure characterizing the monomer displacements is the MSD
$\MSDall(t) \equiv \la (\rvec(t)-\rvec(0))^2 \ra$ displayed in the inset of 
Fig.~\ref{fig_BFMbeads} (with $\rvec(t)$ being the particle position at time $t$).
As one expects, the displacements become uncorrelated for $t \gg 1$, 
i.e. $\MSDall(t) \approx 2d D t$ with $D$ being the monomer self-diffusion constant:
$D=0.0187$ for $d=1^+$, $D=0.0154$ for $d=2^+$, $D=0.0382$ for $d=3$.
The typical displacement $\xi(t) \equiv \MSDall^{1/2}(t)$ thus scales as 
$\xi(t) \sim t^{\alpha}$ with  $\alpha=1/2$.
However, small deviations are clearly visible for short times if $\MSDall(t)/(2d D t)$ is plotted 
in log-linear coordinates. The deviations are particulary strong for $d=1^+$.
The effective random forces acting on the beads are thus not completely white.
Obviously, one might try to characterize these deviations by fitting various polynomials to the
measured MSD. Since one needs to substract, however, the huge free diffusion contribution from
the measured signal to obtain tiny deviations this is a numerical difficult if not impossible route.

In analogy to the static angular correlation function $\Ps \sim \partial_s^2 \Rs^2$ 
allowing to make manifest deviations from the Gaussian chain assumption, 
it is numerically much better to {\em directly} compute the second derivative 
of $\MSDall(t)$ with respect to time. How this can be done 
is illustrated in the main panel of Fig.~\ref{fig_BFMbeads}.
We sample equidistant series of configurations at time intervals $\deltat = 1, 10, 100, \ldots$ 
as indicated by the open symbols.
Each time series contains $10^4$ configurations. Averaging over all possible pairs of configurations
$(t_0,t_0+t)$ we compute the displacement correlation function
$C(t,\deltat) \equiv \la \uvec(t_0+t) \cdot \uvec(t_0) \ra_{t_0}/\deltat^2$,
i.e. a four-point correlation function of the monomer trajectories
with $\uvec(t) = \rvec(t+\deltat) - \rvec(t)$ being the monomer displacement vector 
at time $t$ in a time interval $\deltat$.
By construction $C(t,\deltat) \equiv 0$ if both displacement vectors are uncorrelated.
Note that if one computed times $t$ shorter than $\deltat$, both displacement
vectors would become trivially correlated, since they describe in part the same particle trajectory.
Hence, $C(t=0,\deltat) = \MSDall(\deltat)/\deltat^2$ and $C(t < \deltat,\deltat) > 0$ 
(not shown). More importantly, for times $t \gg \deltat$ one has
\begin{equation}
C(t,\deltat) = \frac{\MSDall(t+\deltat) -2 \MSDall(t) + \MSDall(t-\deltat) }{2 \deltat^2}
 \approx \frac{1}{2} \partial_t^2 \MSDall(t) \ \deltat^0 
\label{eq_Ctht}
\end{equation}
as one readily sees by applying finite-difference operators with respect to time to the monomer MSD.
As can be seen for $d=1^+$, the $\deltat$-dependence drops indeed out for $t/\deltat > 1$. 
We thus often avoid the second index $\deltat$ and write $C(t)$ for the displacement correlation function.
Obviously, the statistics deteriorates for large $t/\deltat$ where fewer 
configuration pairs contribute to the average (taking apart that the signal itself decays). 
It is for this reason that we need a hierarchy of time series of different $\deltat$.
Taking for each time series only the first decade of data ($2 \le t/\deltat < 20$),
these reduced data sets are pasted together and averaged logarithmically.
These logarithmic cummulants are given for each dimension and compared to the exponents 
$\omega=3/2$ (thin line), $\omega=2$ (dashed line) and $\omega=5/2$ (bold line)
predicted by Eq.~(\ref{eq_VCFcolloid}) for $d=1$, $d=2$ and $d=3$, respectively.
The data agrees over more than two orders of magnitude in time with the prediction,
especially for $d=1^+$. 
If one is satisfied with less orders of magnitude it is sufficient to check the exponents 
using just a time window $\deltat=1$ 
as can be seen for the open spheres.
The superposition of data from time series with different $\deltat$ is just a numerical trick
which reduces the number of configurations to be stored and the number of configuration pairs
to be computed for a given time $t$.

\subsection{Polymer melts without topological constraints}
\label{sub_comp_jean}
\begin{figure}[tb]
\centerline{\resizebox{0.6\columnwidth}{!}{\includegraphics*{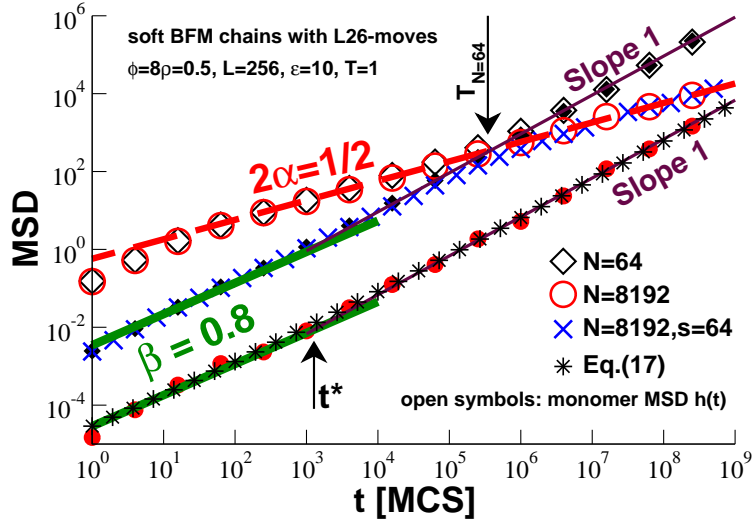}}}
\caption{
Various MSD for BFM chains of lengths $N=64$ and $N=8192$.
The open symbols refer to the monomer MSD $\MSDmon(t)$,
the filled symbols to the MSD $\MSDcmN(t)$ of the COM of chains,
the crosses to the MSD $\MSDcms(t)$ of the COM of subchains of arc-length $s=64$ 
of total chains of length $N=8192$. 
The dashed line indicates the monomer MSD expected for Rouse chains for $t \ll \TN$,
the thin solid line the free diffusion limit expected for uncorrelated random forces.
The bold lines represent the effective exponent $\beta=0.8$, Eq.~(\ref{eq_betadef}).
The stars show Eq.~(\ref{eq_Ct2MSD}),
i.e. the time window $t \ll \tstar$ is described by $\beta=3/4$.
\label{fig_MSD_jean}
}
\end{figure}

\subsubsection{Mean-square displacements}
\label{par_comp_jean_MSD}

Having shown that scale-free dynamical correlations exist for dense BFM beads as expected for 
overdamped colloids (Fig.~\ref{fig_dipole}), we turn now our attention to 3D melts of long and 
flexible homopolymers. We focus here on systems with finite monomer overlap penalty $\overlap=10$ 
and volume fraction $\phi=0.5$.
Due to the use of the finite  overlap penalty and the topology non-conserving local L26-moves,
the dynamics is indeed of Rouse-type as can be seen from the various MSD presented 
in Fig.~\ref{fig_MSD_jean} for chains of length $N=64$ (diamonds) and $N=8192$ (spheres). 

The monomer MSD $\MSDmon(t)$ is indicated by the open symbols.
Note that for the large chain lengths sampled here it is inessential whether this average
is performed over all monomers (as in Sec.~\ref{sub_comp_beads}) or only over a few monomers
in the (curvilinear) center of chains at $n \approx N/2$ as in Refs.~\cite{Paul91a,Paul91b,KBMB01}.
As expected from the Rouse model we obtain for short times
\cite{DoiEdwardsBook}
\begin{equation}
\MSDmon(t) = b^2 (W t)^{1/2} N^0 \mbox{ for }  10 \ll t \ll \TN \approx N^2/W
\label{eq_MSDmonshort}
\end{equation}
as indicated by the dashed line, i.e. the typical displacement increases as $\xi(t)\sim t^{\alpha}$
with $\alpha=1/4$. 
As can be seen for $N=64$, the monomers diffuse again freely with a power-law slope $1$ 
for times larger than the Rouse time $\TN$. We remind that it was 
% neither computationally feasible nor 
not the aim of the present work to sample for our larger chains ($N > 1000$) over the huge times
needed to make this free diffusion regime accessible.
Following Paul {\em et al.} \cite{Paul91a,Paul91b}, the short-time power law, 
Eq.~(\ref{eq_MSDmonshort}), can be used to determine the effective monomer mobility. 
We obtain $W(\overlap=10)=0.003$ for our main working point. 
Mobilities for other penalties are listed in Tab.~\ref{tab_overlap}.
As may be seen from Fig.~\ref{fig_AW}, $W(\overlap)$ decays with increasing excluded volume
just as the acceptance rate $A(\overlap)$ (spheres) of the hopping attempts, but the decay 
is even more pronounced for $W(\overlap)$, i.e. increasingly more accepted moves do not 
contribute to the effective motion (``cage effect") \cite{Paul91b}. 

The full symbols displayed in Fig.~\ref{fig_MSD_jean} refer to the MSD 
$\MSDcmN(t)$ of the chain COM defined in Eq.~(\ref{eq_betadef}).
As one expects for Rouse chains, the amplitude of $\MSDcmN(t)$ decreases inversely with $N$
and the diffusion appears to be uncorrelated ($\MSDcmN(t) \sim t$) at least for times 
$t \gg \tstar \approx 10^3 N^0$ as indicated by the vertical arrow.
Obviously, the center-of-mass MSD $\MSDcmN(t)$ and the monomer MSD $\MSDmon(t)$ merge for times
beyond the Rouse time ($t \gg \TN$). Fortunately, since $\MSDcmN(t)$ becomes linear 
for $\tstar \ll \TN$, it is always possible, even for our largest chain length $N$,
to measure the chain self-diffusion coefficient $\DN$ by plotting $\MSDcmN(t)/ 6 t$
in log-linear coordinates. 
For our main working point we thus obtain $N \DN(\overlap=10) \approx 0.009$ for all $N$. 
Values for other $\overlap$ are again given in Tab.~\ref{tab_overlap} and 
are represented in Fig.~\ref{fig_AW} (diamonds). 
That these values are consistent with Rouse dynamics can be checked by comparing the measured
diffusion coefficients with the values obtained using the local mobilities $W$ according to
\cite{DoiEdwardsBook}
\begin{equation}
N \DN  = \frac{\pi}{4} \frac{\be^2}{d} \ W.
\label{eq_W2DN}
\end{equation}
As shown by the dashed line in Fig.~\ref{fig_AW}, Eq.~(\ref{eq_W2DN}) agrees well with the
directly measured diffusion coefficients. Similarly, it is possible (at least for our shorter chains)
to measure the longest Rouse relaxation time $\TN$ by an analysis of the Rouse modes and to compare it 
with $\TN = 4 N^2 / \pi^3 W$
\cite{DoiEdwardsBook}.
We obtain again a nice agreement between directly and indirectly computed relaxation times (not shown).

Up to now we have insisted on the fact that our systems are to leading order of Rouse type
and we have characterized them accordingly. However, deviations from the Rouse picture are clearly
revealed for short times, especially for $\MSDcmN(t)$ in agreement with the
literature \cite{Paul91b,Shaffer95,KBMB01,HT03,Briels02,Paul98,Smith00,PaulGlenn}. 
(That the monomer MSD $\MSDmon(t)$ also deviates for very short times is due to the lattice model,
i.e. to a trivial lower cut-off effect associated to the discretization.)
In agreement with Eq.~(\ref{eq_betadef}), the short-time COM motion can be characterized by an effective 
exponent $\beta \approx 0.8$ (bold lines). Since in our BFM version topological constraints are irrelevant,
this confirms the finding by Shaffer \cite{Shaffer94,Shaffer95} that the deviations found for short chains using 
the classical BFM algorithm with topological constraints \cite{Paul91b} cannot alone be attributed to
precursor effects to reptational dynamics (as discussed in Sec.~\ref{sub_conc_outlook}). 

Before we turn to the more precise numerical characterization of these deviations by means of the
associated displacement correlation function, let us ask whether the observed colored forces 
acting for short times on the COM of the entire chain are also relevant on the scale of subchains of arbitrary 
arc-length $s$ ($1 \ll s \le N$). To answer this question we compute the MSD displacement 
$\MSDcms(t) = \la (\rs(t)-\rs(0))^2 \ra$ associated to the subchain center-of-mass $\rs(t)$ 
as shown in Fig.~\ref{fig_MSD_jean} for subchains of length $s=64$ in the middle of total chains of 
length $N=8192$ (crosses).
Since for short times the subchain does not ``know" that it is connected to the rest of the chain,
one expects it to behave as a total chain of the same length ($s=N$). This is indeed borne out
by our data which are well described by
\begin{equation}
s \MSDcms(t) \approx N \MSDcmN(t) \mbox{ for } t \ll \Ts \approx s^2/W
\label{eq_MSDlocality}
\end{equation} 
for all chain length $N$ and subchain length $s$ studied. The subchains reveal thus for
sufficiently short times the {\em same} colored forces ($\beta\approx 0.8$) as the total chain 
as can be clearly seen from the example given in Fig.~\ref{fig_MSD_jean}.
For larger times, $\Ts \ll t $, the subchain becomes ``aware" that it is connected to the rest of the chain
and gets enslaved by the monomer MSD. We thus observe
\begin{equation}
\MSDcms(t) \approx \MSDmon(t) \approx b^2 (W t)^{1/2} s^0 N^0
\mbox{ for }  \Ts \ll t \ll \TN
\label{eq_MSDcmsintermediate}
\end{equation}
and, obviously, $\MSDcms(t) \approx \MSDcmN(t) \approx \MSDmon(t) \approx 6 \DN t$ for even larger times $t \gg \TN$.

\subsubsection{Locality and relevant exponent $\alpha$}
\label{par_comp_jean_comments}
Two comments are in order here. First, it should be noticed that Eq.~(\ref{eq_MSDlocality}) expresses 
the fact that the effective forces acting on the $N/s$ subchains of length $s$ in a chain of total 
length $N$ add up {\em independently} to the forces acting on the total chain. In this sense 
Eq.~(\ref{eq_MSDlocality}) states that the deviations from the Rouse picture must be {\em local}
\cite{foot_locality}. 
We will explicitly verify this below (Fig.~\ref{fig_VCF_jeansub}).
Second, if one chooses following Eq.~(\ref{eq_Ctht}) 
an arbitrary time window $\deltat$ to characterize the displacement correlations,
this corresponds to dynamical blobs containing $s \approx \sqrt{W \deltat}$ adjacent monomers,
which must move together due to chain connectivity. Eq.~(\ref{eq_MSDcmsintermediate}) implies now that
the dipole field \cite{foot_dipolefield} associated with the COM of these $s$-subchains and 
created at $t=0$ by a tagged $s$-subchain must decay according to a typical displacement 
$\xi(t) \approx \MSDcms^{1/2}(t) \sim t^{\alpha}$ with $\alpha=1/4$.
It is this exponent $\alpha$ which is mentioned in the Introduction, Eq.~(\ref{eq_keyomega}).
Since this exponent is smaller than for freely diffusing colloids ($\alpha=1/2$), 
the dipole field of the COM of the subchains must decay more slowly and one expects, accordingly,
a much weaker decay of the associated VCF. 

\subsubsection{Center-of-mass velocity correlation function}
\label{par_comp_jean_VCF}
\begin{figure}[tb]
\centerline{\resizebox{0.6\columnwidth}{!}{\includegraphics*{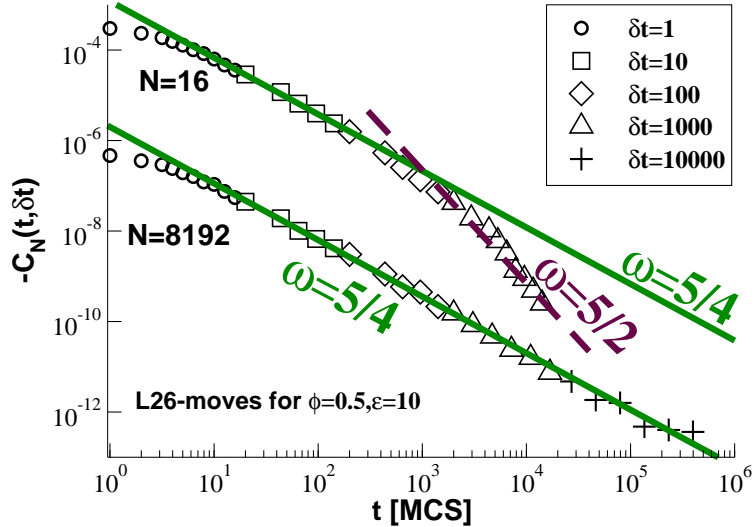}}}
\caption{VCF $\CN(t,\deltat)$ for two chain lengths, $N=16$ (top) and $N=8192$ (bottom), 
obtained using L26-moves. We only indicate for each $\deltat$ the data window which is 
used to construct (by logarithmic averaging) the final VCF $\CN(t)$. 
The bold lines represent the exponent $\omega=5/4$
which is generally observed for times $10 \ll t \ll \TN$.
The exponent $\omega=5/2$ (dashed line) indicated for $N=16$ is expected for $t \gg \TN$.
\label{fig_VCF_jeanconstruct}
}
\end{figure}

\begin{figure}[tb]
\centerline{\resizebox{0.6\columnwidth}{!}{\includegraphics*{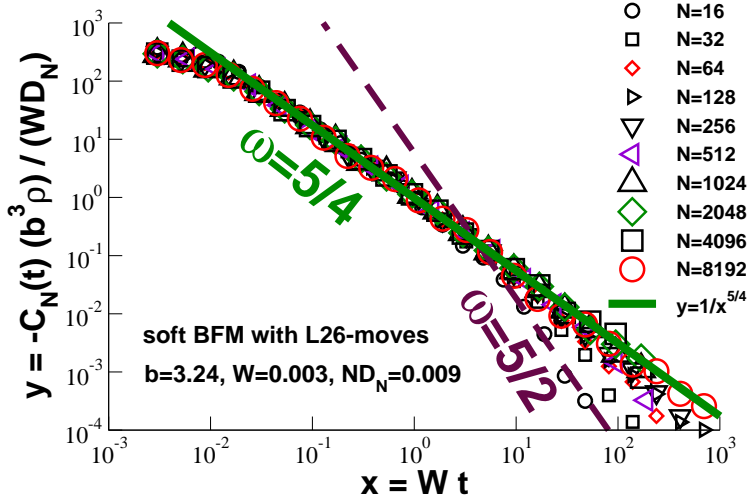}}}
\caption{
Reduced VCF $y = -\CN(t) (b^3 \rho) /(W \DN)$ for different chain lengths $N$ as a function 
of the reduced time $x=W t$ with $W$ being the local mobility,
$\DN \approx 1/N$ the self-diffusion constant. 
For times $t \ll \TN$ the VCF thus scales as $\CN(t) \sim 1/N$,
i.e. the correlations must be due to $\sim N$ local independent events.
The exponent $\omega=5/4$ (bold line) is observed over up to five orders of magnitude in time.
\label{fig_VCF_jean}
}
\end{figure}

\begin{figure}[tb]
\centerline{\resizebox{0.6\columnwidth}{!}{\includegraphics*{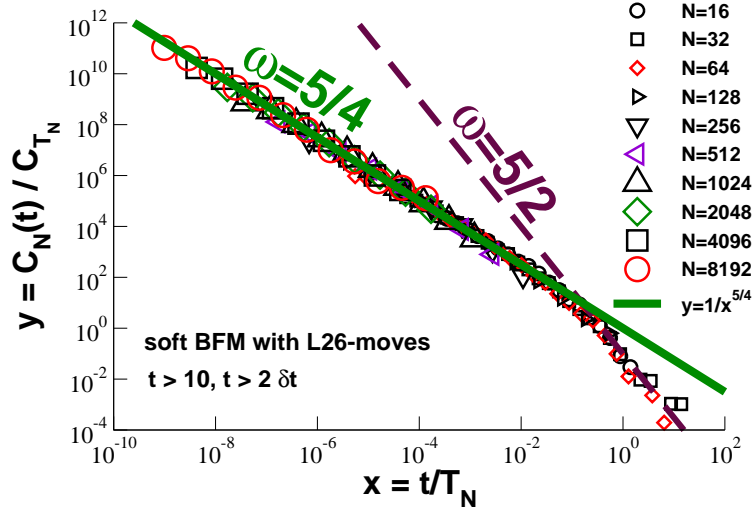}}}
\caption{
Collapse of the reduced VCF $y=\CN(t)/\CNstar$ 
with $\CNstar \sim 1/N^{7/2}$ as a function of $x=t/\TN$ as suggested by Eq.~(\ref{eq_key}). 
The large time behavior ($x \gg 1$) is described
by the exponent $\omega = 5/2$ (dashed line) expected for overdamped colloids, Eq.~(\ref{eq_VCFcolloid}). 
The exponent $\omega = 5/4$ (bold line) for $x \ll 1$ is confirmed over nearly eight
orders of magnitude. This is the central numerical result of this study.
\label{fig_VCF_jeantau}
}
\end{figure}

\begin{figure}[tb]
\centerline{\resizebox{0.6\columnwidth}{!}{\includegraphics*{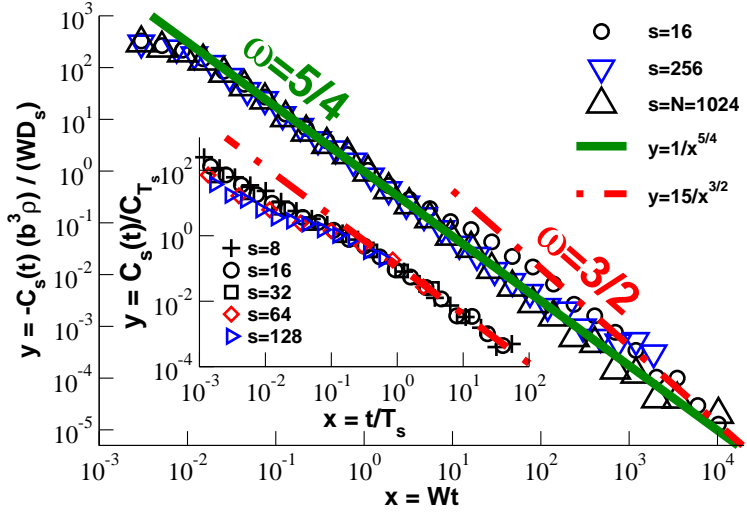}}}
\caption{VCF $\Cs(t)$ for subchains of arc-length $s$ obtained for 
chains of length $N=1024$ using an overlap penalty $\overlap=10$ and L26-moves.
Main panel:
For $t \ll \Ts \sim s^2$ the subchains scale as the total chains,
i.e. $\Cs(t) \sim s^{-1} t^{-5/4}$ (bold line). For intermediate times $\Ts \ll t \ll \TN$
the displacement of subchains ($1 \ll s\ll N$) follows the monomer MSD and we find thus
$\Cs(t) \sim s^0 (b W)^2 (Wt)^{1/2-2}$ (dash-dotted line). 
Inset: Setting $\Ts=s^2/W$ and $\Csstar = - (b W)^2 s^{-3}$ and plotting
$y=\Cs(t)/\Csstar$ as a function of $x=t/\Ts$ the data scales for $x \gg 1$.
The dashed slope corresponds to $y= 1/8x^{3/2}$.
\label{fig_VCF_jeansub}
}
\end{figure}

Following the numerical strategy used in Sec.~\ref{sub_comp_beads} for BFM beads,
we characterize now in more detail the dynamical correlations already visible from the 
chain MSD $\MSDcmN(t)$ and the subchain MSD $\MSDcms(t)$ by computing directly
their second derivative with respect to time, i.e. the displacement correlation functions
$\CN(t) \approx \partial_t^2 \MSDcmN(t)/2$ 
(Figs.~\ref{fig_VCF_jeanconstruct}, \ref{fig_VCF_jean} and \ref{fig_VCF_jeantau}) 
and 
$\Cs(t) \approx \partial_t^2 \MSDcms(t)/2$ (Fig.~\ref{fig_VCF_jeansub}).

The displacement correlation function $\CN(t,\deltat) = \la \uvec(t+t_0) \cdot \uvec(t_0) \ra_{t_0}/\deltat^2$ 
for the displacement vector  $\uvec(t) = \rN(t+\deltat)-\rN(t)$ of the chain COM $\rN(t)$ is shown 
in Fig.~\ref{fig_VCF_jeanconstruct} for two chain lengths, $N=16$ (top) and $N=8192$ (bottom). 
Averages are again performed over all configuration pairs $(t_0,t+t_0)$ possible in the set of $10^4$
configurations sampled for each $\deltat$ indicated.
As in Sec.~\ref{sub_comp_beads} we find that $\CN(t,\deltat) \sim \deltat^0$ for $t \gg \deltat$.
For clarity, only the data subset is indicated which is used to construct the 
cummulated final VCF $\CN(t)$ (as shown below in Fig.~\ref{fig_VCF_jean}). 
The bold lines represent the predicted short-time exponent $\omega=5/4$.
We emphasize that this exponent can be observed for $N=8192$ over nearly five orders of magnitude.
For $N=16$ we also indicate the exponent $\omega=5/2$ expected for times $t \gg \TN$ 
where the chains should behave as colloids according to Eq.~(\ref{eq_VCFcolloid}).
The magnitude of the signal decreases strongly with $N$,
which together with the fact that fewer chains per box are available makes the determination of 
the VCF more difficult with increasing chain length.

The $N$-dependence of the VCF for $t \ll \TN$ is further analysed in Fig.~\ref{fig_VCF_jean}
where we plot the reduced VCF $y=-\CN(t) (b^3 \rho) / (W \DN)$ {\em vs.} the reduced time $x=Wt$ using 
the effective monomer mobility $W$ and the diffusion coefficient $\DN$ determined above. 
This scaling makes the axes dimensionless and rescales the vertical axis by a factor $N$.
As shown by the successful data collapse for chain lengths ranging from $N=16$ up to $N=8192$
on the power-law slope indicated by the bold line, the VCF scales exactly as $\CN(t) \sim 1/N$ 
for $t \ll \TN$. This confirms the already stated ``locality" of the correlations \cite{foot_locality}, 
Eq.~(\ref{eq_MSDlocality}), i.e. the effective forces acting on subchains add up 
independently to the forces acting on the entire chain.
We have still to motivate the precise form used for the rescaling of the axes.
According to Eq.~(\ref{eq_key}) we claim that the VCF $\CN(t)$ scales as a function of 
the reduced time $t/\TN$. Substituting the typical chain size $\RN \approx b N^{1/2}$ and the chain
relaxation time $\TN \approx \RN^2/\DN \approx N^2/W$ and using $\omega=5/4$ this yields
\begin{equation}
y \equiv - \be^3 \rho \ \frac{\CN(t)}{W \DN} = c \ (W t)^{-5/4} N^0
\mbox{ for } t \ll \TN
\label{eq_VCF_Wt}
\end{equation}
with $c$ being an empirical dimensionless constant. 
The bold slope indicated in the plot corresponds to a value $c \approx 1$.
Interestingly, since $\CN(t) \approx \partial_t^2 \MSDcmN(t)/2$,
it follows from Eq.~(\ref{eq_VCF_Wt}) that 
\begin{equation}
\MSDcmN(t) = 6 \DN t \left( 1 + \frac{16 c}{9 \be^3\rho} (Wt)^{-1/4} \right).
\label{eq_Ct2MSD}
\end{equation}
As may be seen from Fig.~\ref{fig_MSD_jean} (stars) for $N=8192$, Eq.~(\ref{eq_Ct2MSD}) with $c=1$
provides an excellent fit of the measured $\MSDcmN(t)$. 
We also note that the second term in Eq.~(\ref{eq_Ct2MSD}) 
dominates the dynamics for $t \ll \tstar \equiv W^{-1} (16 c / 9 b^3\rho)^4 \ N^0 \approx 10^3$.
This is indicated by the vertical arrow in Fig.~\ref{fig_MSD_jean}. Hence, for $t \ll \tstar$
the stars correspond to an exponent $\beta = 2- \omega= 3/4$, Eq.~(\ref{eq_omega2beta}),
being close to the phenomenological exponent $\beta \approx 0.8$ which motivated our study.
The central advantage of computing the VCF $\CN(t)$ lies in the fact that it allows us thus to make 
manifest that (negative algebraic) deviations from the Rouse behavior exist for {\em all} times and 
not just for $t \ll \tstar$. 

Returning to our discussion of Fig.~\ref{fig_VCF_jean} we emphasize that the VCF of shorter chains 
decays more rapidly for large times following roughly the dashed line corresponding to 
the exponent $\omega=5/2$ expected for effective colloids. 
We verify in Fig.~\ref{fig_VCF_jeantau} that this bending down of the VCF is described 
by the announced scaling in terms of a reduced time $x=t/\TN$ 
and a vertical axis $y= \CN(t)/\CNstar$ using the amplitude
$\CNstar \equiv \CN(t=\TN) \approx - (\RN/\TN)^2 \rhostar/\rho \sim -1/N^{7/2}$ stated in Eq.~(\ref{eq_key}).
As shown by the data collapse, there is only one 
relevant time scale in this problem, namely the chain relaxation time $\TN$, 
for which the deviations
for short times, where polymer physics matters ($\omega=5/4$), and for large times,
where polymer chains behave as effective colloids ($\omega=5/2$), nicely match.
It is worthwhile to emphasize that Eq.~(\ref{eq_key}) together
with the locality of the deviations, $\CN(t) \sim 1/N$, immediately
imply the exponent $\omega$. This can be seen by counting the powers
of the chain length $N$,
\begin{equation}
-1 \stackrel{!}{=} (1/2-2) 2 + (1 - d/2) + 2 \omega, 
\label{eq_Ncounting}
\end{equation}
which implies $\omega = (d+ 2)/4 = 5/4$ in agreement with Eq.~(\ref{eq_keyomega})
and the numerically observed time dependence.
Assuming Eq.~(\ref{eq_key}), the exponents for $N$ and $t$ thus contain the same information.

The subchain VCF $\Cs(t)$ presented in Fig.~\ref{fig_VCF_jeansub} has been obtained as the total
chain VCF $\CN(t)$, the only difference being that the COM $\rs(t)$ of the subchain
defines now the displacement vector $\uvec(t) = \rs(t+\deltat)-\rs(t)$.
The data presented in the main panel is rescaled as in Fig.~\ref{fig_VCF_jean}
with $\Ds = 0.009/s$ setting now the diffusion constant.
For small times $t \ll \Ts \approx s^2/W$, all data sets
collapse on the same slope with exponent $\omega=5/4$ (bold line)
as in Eq.~(\ref{eq_VCF_Wt}). This confirms that 
\begin{equation}
\frac{\Cs(t)}{W\Ds} \approx \frac{\CN(t)}{W\DN} \sim -(W t)^{-5/4} \mbox{ for } t \ll \Ts \approx s^2/W
\label{eq_VCFlocality}
\end{equation}
in agreement with Eq.~(\ref{eq_MSDlocality}).
This shows that the same deviations occur for arbitrary subchains
and that the colored forces acting on the subchains add up independently to the
effective forces acting on the total chain.
Since for intermediate times $\Ts \ll t \ll \TN$ the subchains are enslaved
by the monomer motion according to Eq.~(\ref{eq_MSDcmsintermediate}), 
it follows using Eq.~(\ref{eq_Ctht}) that $\Cs(t) \sim s^0 (b W)^2 (Wt)^{1/2-2}$
(dash-dotted line). This scaling can be better seen in the inset of Fig.~\ref{fig_VCF_jeansub}
where we plot $y=\Cs(t)/\Csstar$ as a function of $x=t/\Ts$ setting
$\Ts \equiv s^2/W$ and $\Csstar \equiv - (b W)^2 s^{-3}$. This allows to bring to 
a nice collapse the subchain VCF for $x \gg 1$. The dash-dotted line corresponds 
to the expected decay $y= 1/8x^{3/2}$ for intermediate times.

\subsection{Robustness of scaling behavior}
\label{sub_comp_robust}
\begin{figure}[tb]
\centerline{\resizebox{0.6\columnwidth}{!}{\includegraphics*{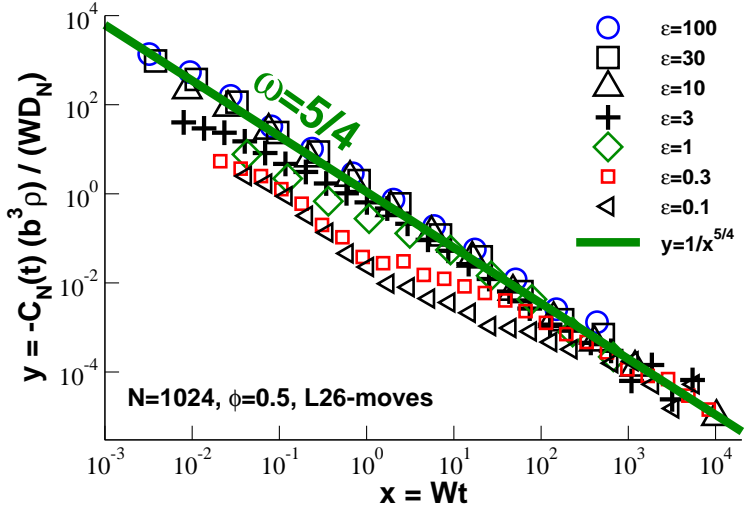}}}
\caption{VCF $\CN(t)$ for various monomer overlap penalties $\overlap$
for chain length $N=1024$, volume fraction $\phi=0.5$ and L26-moves.
For sufficiently large times ($t \ll \TN$) all data appears to collapse
on the same asymptotic exponent $\omega=5/4$ (bold line) for 
incompressible polymer solutions. Additional physics is visible at short times. 
\label{fig_VCF_overlap}
}
\end{figure}

We have focused above on one specific choice of operational parameters. 
In order to check whether our key predictions Eq.~(\ref{eq_key}) and (\ref{eq_keyomega}) 
remain valid more generally under different conditions, additional simulations have 
been performed which we summarize here:

{\em (i)}
Switching to the L06-moves of the original BFM algorithm, 
but keeping the same static conditions ($\phi=0.5$, $\overlap=10$),
the same scaling is obtained for the MSDs and the VCFs as before (not shown).
Using the directly measured local monomer mobility $W=0.002$ and the self-diffusion coefficient 
$\DN = 0.007/N$ we get for the VCF exactly the same scaling as in Fig.~\ref{fig_VCF_jean}. 
This is consistent with the fact that Eq.~(\ref{eq_key}) does not depend explicitly on the 
specific local dynamics.

{\em (ii)}
Using again L26-moves we have varied the overlap penalty $\overlap$ at constant chain length 
$N=1024$ and volume fraction $\phi=0.5$. Since Eq.~(\ref{eq_key}) does only depend {\em implicitly} on $\gT$,
this suggest that Eq.~(\ref{eq_VCF_Wt}) for times $t \ll \TN$ should also remain valid if one uses for rescaling 
of the axes the measured effective bond length $b(\overlap)$, the mobility $W(\overlap)$ and the diffusion 
constant $\DN(\overlap)$ indicated in Tab.~\ref{tab_overlap}.
This is in fact borne out for the asymptotic behavior of the rescaled VCF displayed in Fig.~\ref{fig_VCF_overlap}.
In analogy to the angular correlation function $\Ps$ for soft melts presented in panel (c) of 
Fig.~\ref{fig_static}, the deviations from Eq.~(\ref{eq_VCF_Wt}) for short times are expected 
since the incompressibility constraint is only felt if distances corresponding to the static 
screening length are probed. 

{\em (iii)} 
A similar plot has been obtained for $\CN(t)$ if we change the dimensionless compressibility by varying 
the volume fraction $\phi$ at constant chain length $N=1024$ and overlap penalty $\overlap=10$ (not shown).  
As long as the chains remain sufficiently entangled, all data merge for large times on the asymptotic behavior 
of incompressible melts, Eq.~(\ref{eq_VCF_Wt}), for $t\ll \TN$. 

%%%%%%%%%%%%%%%%%%%%%%%%%%%%%%%%%%%%%%%%%%%
\section{Conclusion}
\label{sec_conc}
\subsection{Summary}
\label{sub_conc_summary}
The incompressibility constraint of polymer melts is known to restrict the fluctuations 
of (sub)chains and generates thus scale-free {\em static} deviations from the Gaussian chain 
statistics \cite{WMBJOMMS04,WBM07,WCK09}.
In this paper we addressed the question of whether the incompressibility constraint 
also causes {\em dynamical} correlations of the (sub)chain displacements. 
To avoid additional correlations we have supposed that the dynamics is locally perfectly overdamped 
(``no momentum conservation") and that the chains may freely intercept (``no reptation"), 
i.e. the relaxational dynamics is assumed to be to leading order of Rouse type \cite{DoiEdwardsBook}.
We have shown that the above conditions may be realized computationally by a variant of the BFM
using finite monomer excluded volume penalties 
and local topology non-conserving MC moves (Fig.~\ref{fig_algo}).
Sampling chain lengths up to $N=8192$ allowed us to carefully check the $N$-scaling of the
deviations (Figs.~\ref{fig_VCF_jean}, \ref{fig_VCF_jeantau}).
Such deviations are visible from the short-time scaling of the COM MSD $\MSDcmN(t)$
presented in Fig.~\ref{fig_MSD_jean} 
confirming published computational and experimental work 
\cite{Paul91b,Shaffer94,Shaffer95,KBMB01,Briels02,HT03,Paul98,Smith00,PaulGlenn}.
We have shown that a better characterization of these deviations 
can be achieved by means of the COM displacement auto-correlation function
$\CN(t) \approx \partial^2_t \MSDcmN(t)/2$. Computing directly the curvature of $\MSDcmN(t)$
this removes the large free diffusion contribution to the chain motion and allows to focus on 
the correlated random forces 
which cause the second term contributing to $\MSDcmN(t)$ according to Eq.~(\ref{eq_Ct2MSD}). 
How such a ``velocity correlation function" can be computed within a MC scheme
has been first illustrated for dense BFM beads (Fig.~\ref{fig_BFMbeads}) confirming the negative 
algebraic decay expected for overdamped colloids, Eq.~(\ref{eq_VCFcolloid}).  
The observed exponent $\omega=(d+2) \alpha$ is expected \cite{DhontBook} due to the coupling of 
a tagged colloid to the gradient of the collective density dipole field (Fig.~\ref{fig_dipole}) 
decaying in time by free diffusion ($\alpha=1/2$).
As shown in Fig.~\ref{fig_VCF_jeantau}, the same exponents (dashed lines) characterize the motion of 
polymer chains for large times ($t \gg \TN$) where the chains behave as effective colloids.
More importantly, we have demonstrated that the observed short-time deviations 
for $\MSDcmN(t)$, Eq.~(\ref{eq_betadef}), can be traced back to the negative analytic decay 
of the correlation function, $\CN(t) \sim - N^{-1} t^{-\omega}$ for $t \ll \TN$ 
with $\omega \approx 5/4$ in $d=3$ (Fig.~\ref{fig_VCF_jean}) 
in agreement with Eq.~(\ref{eq_keyomega}) \cite{foot_capillary}. 
That the correlation functions decays inversely with mass shows that the process is {\em local}
\cite{foot_locality}, i.e. the displacement correlations of subchains add up independently 
(Fig.~\ref{fig_VCF_jeansub}).
Assuming according to the key scaling relation Eq.~(\ref{eq_key})
the relaxation time $\TN$ to be the only characteristic time, both 
asymptotic regimes can be brought to a data collapse (Fig.~\ref{fig_VCF_jeantau}).
The short-time exponent $\omega=(d+2)/4$ is implied by Eq.~(\ref{eq_key})
and the locality of the correlations, Eq.~(\ref{eq_Ncounting}).
A deeper insight is obtained by generalizing the well-known displacement correlations 
of overdamped colloids (Fig.~\ref{fig_dipole}) to the displacement field of subchains 
of length $s \sim \deltat^{1/2}$ with $\deltat$ being the time window used to define 
the displacements. Since subchains repel each other due to the incompressibility constraint,
a tagged subchain is pulled back to its original position by the subchain dipole field
\cite{foot_dipolefield}.
Since for times $\deltat \ll t \ll \TN$ the relevant dipole field decays much slower 
than for colloids ($\alpha =1/2 \to 1/4$), the correlations are much more pronounced.

\subsection{Outlook}
\label{sub_conc_outlook}
In this study we have deliberately tuned our model to avoid topological constraints.
Obviously, these constraints are expected to matter for the dynamics of real polymer 
melts \cite{DegennesBook,DoiEdwardsBook}. It is thus of interest
to see how the presented picture changes if topology is again switched on by
using the topology conserving L06-moves of the classical BFM algorithm. 
\begin{figure}[tb]
\centerline{\resizebox{0.6\columnwidth}{!}{\includegraphics*{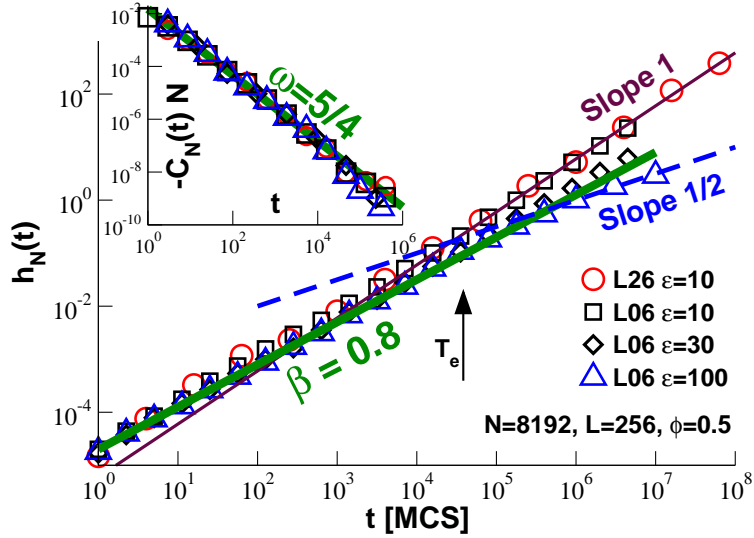}}}
\caption{Topology induced additional correlations in BFM melts for chain length $N=8192$.
Main panel: MSD $\MSDcmN(t)$ {\em vs.} time $t$ comparing L26-moves for $\overlap=10$ (spheres) 
to L06-moves for different overlap penalties $\overlap$.
Topological constraints become relevant for L06-moves with $\overlap \gg 10$.
Inset: The reduced VCF $-\CN(t) N$ confirms that for all system classes
the short-time dynamics is described by the {\em same} exponent $\omega=5/4$ (bold line).
\label{fig_rept}
}
\end{figure}
Preliminary data comparing systems of chain length $N=8192$
is presented in Fig.~\ref{fig_rept}. The dynamics is still of Rouse-type at 
small overlap penalties as can be seen for $\overlap=10$.
With increasing $\overlap$ the crossing of the chains gets more improbable 
and the topological constraints become more relevant. This can be seen for 
$\overlap=30$ (diamonds) and even more for $\overlap=100$ (triangles). 
The latter data set clearly approaches the power-law slope $1/2$ (dashed line) 
expected from reptation theory for times larger than the entanglement time $\Te$.
The vertical arrow indicates a value for this time obtained from an analysis of $\MSDmon(t)$. 
Apart from the much larger chain length used, our data for $\overlap=100$ is 
consistent with the results obtained using the classical BFM  
with $\overlap=\infty$ \cite{Paul91b,KBMB01}. 
As in Fig.~\ref{fig_MSD_jean} the bold line represents the effective exponent $\beta = 0.8$, Eq.~(\ref{eq_betadef}).  
Superficially, it does a better job for systems with conserved topology due to the broad 
crossover to the entangled regime.
However, this apparent exponent is by no means deep as revealed by the displacement
correlation function $\CN(t)$ plotted in the inset. For short times {\em all} systems
are well-described by the {\em same} exponent $\omega=5/4$ (bold line) 
in agreement with Eq.~(\ref{eq_keyomega}). 
Note that the final points given for $\overlap=100$ decay slighly more rapidly.
Unfortunately, neither the length of the analyzed time series nor the precision 
of our data does currently allow us to show that 
\begin{equation}
\CN(t) \approx - \frac{1}{N} \left(\frac{\de}{\Te} \right)^2 \left( \frac{t}{\Te} \right)^{-3/2} 
\mbox{ for } t \gg \Te
\label{eq_CNentangled}
\end{equation}
as expected for reptating chains with $\de \sim \Te^{1/2}$ being the tube 
diameter \cite{DegennesBook}. 
Much longer time series are currently under production to clarify this issue.
The numerical demonstration is obviously challenging, since the difference between
the exponents $3/2$ and $5/4$ is rather small.
In any case it is thus due to the dynamical correlations 
first seen in the BFM simulations of Paul {\em et al.} \cite{Paul91b,PaulGlenn} 
that the crossover between Rouse and reptation regimes becomes broader and
more difficult to describe than suggested by the standard Rouse-reptation theory 
\cite{DegennesBook,DoiEdwardsBook} which 
does not take into account the (static and dynamical) correlations 
of the composition fluctuations imposed by the incompressibility constraint 
\cite{foot_capillary,foot_related}.

%%%%%%%%%%%%%%%%%%%%%%%%%%%%%%%%%%%%%%%%%%%
\begin{acknowledgments}
We thank S.P.~Obukhov (Gainesville) and A.N.~Semenov (Strasbourg) for helpful discussions.
A.C. acknowledges the MIUR (Italian Ministry of Research) for support within the 
program ``Incentivazione alla mobilit\`a di studiosi stranieri e italiani residenti all'estero",
and P.P. a grant by the IRTG ``Soft Matter Science".
\end{acknowledgments}

%%%%%%%%%%%%%%%%%%%%%%%%%%%%%%%%%%%%%%%%%%%
%%%%%%%%%%%%%%%%%%%%%%%%%%%%%%%%%%%%%%%%%%%

%%%\input{bibl}
%\bibliography{../bibl,foot,../../bibl/BOOK,../../bibl/JPW,../../bibl/ANS,../../bibl/deGennes,../../bibl/Obukhov,../../revIntra/biblBFM,../../bibl/KK}

\begin{thebibliography}{39}
\bibitem[{foo({\natexlab{a}})}]{foot_Alder}
\bibinfo{note}{The discussion is simplified. Strictly speaking, it is not the
  particle diffusion which sets the dynamical length scale but the diffusion of
  the transverse momentum \cite{HansenBook}.}
\bibitem[{foo({\natexlab{b}})}]{foot_colloid}
\bibinfo{note}{All prefactors are omitted for simplicity. Especially, we do not
  distinguish between the {\em self} diffusion of the test particle and the
  {\em collective} diffusion of the field which may be characterized by rather
  different diffusion constants.}
\bibitem[{foo({\natexlab{c}})}]{foot_MC}
\bibinfo{note}{In the context of MC simulations ``velocity" refers
  strictly speaking to the displacement $\uvec(t,\deltat) = \rvec(t+\deltat) -
  \rvec(t)$ per time increment $\deltat$. The VCF $C(t) \equiv \la
  \uvec(t,\deltat) \cdot \uvec(0,\deltat) \ra/ \deltat^2$, measuring for $t \gg
  \deltat$ the second derivative of the mean-square displacement $\MSDmon(t)$
  with respect to time, may be called more precisely ``displacement correlation
  function". See Sec.~\ref{sub_comp_beads} for details.}
\bibitem[{foo({\natexlab{d}})}]{foot_Psdef}
\bibinfo{note}{The angular correlation function $\Ps$ may be defined as $\la
  \lvec_{n} \cdot \lvec_{m=n+s} \ra_n / \la \lvec_n^2 \ra $ where $\lvec_n$
  stands for the bond vector connecting two adjacent monomers $n$ and $n+1$
  \cite{WMBJOMMS04}.}
\bibitem[{foo({\natexlab{e}})}]{foot_rhostartho}
\bibinfo{note}{That this factor arises may be seen from the large-time limit
  where the polymer behave as effective colloids repelling each other with a
  penalty $\Ustar(N) \approx \rhostar(N)/\rho$.}
\bibitem[{foo({\natexlab{f}})}]{foot_capillary}
\bibinfo{note}{We have also preformed BFM simulations of polymer melts confined
  to effectively 2D thin slits and effectively 1D capillaries. In agreement
  with Eq.~(\ref{eq_keyomega}) these simulations show that $\omega \approx 1$
  for slits and $\omega \approx 3/4$ for capillaries.}
\bibitem[{foo({\natexlab{g}})}]{foot_locality}
\bibinfo{note}{The ``locality" of the correlations described by
  Eq.~(\ref{eq_MSDlocality}) or Eq.~(\ref{eq_VCFlocality}) does not imply that
  the displacements of subchains around the reference subchain are
  $\delta(r)$-correlated.}
\bibitem[{foo({\natexlab{i}})}]{foot_related}
\bibinfo{note}{We have checked by MD simulations of a standard bead-spring
  model that $\CN(t) \approx N^{-1} t^{-5/4}$ for $t \ll \TN$ holds as long as
  a large friction constant $\gamma$ is used for the Langevin thermostat
  \cite{ANS11a}. Much stronger and non-local dynamical correlations are
  revealed, however, for $\gamma \to 0$. These numerical results will be
  discussed elsewhere together with a broader theoretical analysis which
  contains the mechanism presented in this paper as a special limit.}
\end{thebibliography}

\end{document}